\documentstyle[aps,twocolumn]{revtex}
\input psfig
\begin{document}
\draft
\title{Fluctuations of Conductance Peak Spacings in
the Coulomb Blockade Regime: Role of Electron-Electron Interaction.}
\author{Ya.~M.~Blanter$^{a,b,*}$,
A.~D.~Mirlin$^{a,c}$, and B.~A.~Muzykantskii$^d$} 
\address{$^a$ Institut f\"ur Theorie der Kondensierten Materie,
Universit\"at Karlsruhe, 76128 Karlsruhe, Germany\\
$^b$ Department of Theoretical Physics, Moscow Institute for Steel and
Alloys, Leninskii Pr. 4, 117936 Moscow, Russia\\
$^c$ Petersburg Nuclear Physics Institute, 188350 Gatchina,
St. Petersburg, Russia\\
$^d$ Department of Physics, University of Warwick, CV4 7AL Coventry,
UK} 
\date{\today}
\maketitle 
\tighten
\begin{abstract}
We study influence of electron-electron interaction on statistics of
Coulomb blockade peak spacings in disordered quantum dots.
It is shown
that the interaction combined with fluctuations of eigenfunctions of
the Fermi sea, enhances the peak spacing fluctuations, in accordance
with recent experiments. In addition, account of the spin degrees
of freedom leads to a pronounced odd-even structure for weak
interaction ($e^2/\epsilon \ll v_F$); in the opposite case
($e^2/\epsilon \gtrsim v_F$) this structure is washed out.
\end{abstract}

Recent experimental studies of chaotic quantum dots in the Coulomb
blockade regime showed unusually large fluctuations of conductance
peak spacings. The r.~m.~s fluctuation of the spacing between
consecutive peaks was found to exceed the mean level spacing $\Delta$
\cite{Marcus,Sivan1}. These fluctuations can not be explained
from the point of view of the standard ``charging energy'' model
(CEM) \cite{CIT} which assumes that the spacing between the
consecutive peaks is equal to 
$S_N = e^2/C + \delta \xi_N$.
Here $C$ is the capacitance of the quantum dot, while $\delta \xi_N$
is the distance between the last filled single-particle level and the
previous one. Since the charging energy $e^2/C$ is a non-fluctuating
quantity in CEM, the peak spacing fluctuations are solely due to the
single-particle energies. Assuming the applicability of the random
matrix theory (RMT) one obtains \cite{Mehta} (we consider the case of
broken time reversal symmetry throughout the paper) 
\begin{equation} \label{rmt}
\mbox{r.~m.~s.} (S_N) = \Delta (3\pi/8 - 1)^{1/2},
\end{equation}
that is less than observed experimentally. 

Below we study the fluctuations of peak spacings in the Coulomb
blockade regime. First we investigate the Coulomb interaction in the
finite system, and find corrections to CEM. Then we demonstrate that
these corrections, being combined with fluctuations of eigenfunctions,
lead to enhancement of spacing fluctuations. This becomes
important for a sufficiently strong interaction and may account for
experimentally observable effects. In addition, we study the effect of
spin degree of freedom and find a pronounced
odd-even structure of conductance peaks, which for strong interactions
is washed out. 

Recently Sivan et al \cite{Sivan1} made an attempt to
explain the large fluctuations of $S_N$ observed in experiment using
numerical diagonalization of a Coulomb system with $N \sim 10$
electrons. However, they considered the range of very strong
interaction $e^2/C > E_F$, where $E_F$ is the Fermi energy, while in
the experiment \cite{Marcus} $e^2/C\sim 0.6 meV$ and $E_F\sim 12 meV$.
Effect of spin has been also recently studied by
Prus et al \cite{Prus} in terms of a disordered Hubbard model.

We consider a diffusive quantum dot ($R \gg l$, $R$ and $l$ being the
size of the dot and the mean free path, respectively) with a large
number of electrons. Most of them are compensated by a positive
background with the density $\bar \rho$; in addition, there is a
number $N$ of excess (uncompensated) electrons, so that the total
charge of the dot is $Ne$. The Hamiltonian of the system is 
\begin{eqnarray} \label{int1}
\hat H &=& \sum_{\lambda} \epsilon_{\lambda} a^{+}_{\lambda}
a_{\lambda} + \frac{1}{2} \sum_{\lambda_i} a^{+}_{\lambda_1}
a^{+}_{\lambda_2} U^{\lambda_1,\lambda_2}_{\lambda_3,\lambda_4}
a_{\lambda_3} a_{\lambda_4}, \\
 \label{int2}
U^{\lambda_1,\lambda_2}_{\lambda_3,\lambda_4} &=& \int d\bbox{r}_1
d\bbox{r}_2 \psi^{*}_{\lambda_1}(\bbox{r}_1)
\psi^{*}_{\lambda_2}(\bbox{r}_2) U(\bbox{r}_1 - \bbox{r}_2)\nonumber \\
&\times& \psi_{\lambda_3}(\bbox{r}_2) \psi_{\lambda_4}(\bbox{r}_1).
\end{eqnarray}
Here $\lambda$ labels eigenstates of non-interacting system with
eigenfunctions $\psi_{\lambda}(\bbox{r})$ and energies
$\epsilon_\lambda$; the correlations of these states have been
calculated in Ref. \cite{BM1}. Furthermore, $U(r) = U_0(\bbox{r})
\equiv e^2/\epsilon r$ is the
 Coulomb interaction, $\epsilon$ being the dielectric
constant. The interaction with  positive
background should be added to Eq. (\ref{int1}). 

{\bf Random-phase approximation in restricted geometry}. As is
well known, it is not sufficient to consider the Coulomb interaction
in the first order of perturbation theory due to its long-range
nature. A common way to improve it is the random phase
approximation (RPA). Whereas it is trivial to solve the RPA equation
in infinite system, restricted geometry makes the situation
more complicated (cf \cite{Buett}). 
In the 3D case the RPA equation for the effective potential $U$ is
\begin{eqnarray} \label{rpa1}
U(\bbox{r}, \bbox{r'}) & = & U_0 (\bbox{r} - \bbox{r'}) - \int d\bbox{r_1}
d\bbox{r_2} U_0 (\bbox{r} - \bbox{r_1}) \nonumber \\
& \times & K(\bbox{r_1},\bbox{r_2}) U(\bbox{r_2}, \bbox{r'}),  
\end{eqnarray}
where the polarization operator $K$ has the form 
\begin{eqnarray*} 
& & K(\bbox{r},\bbox{r'}) \approx \nu \left(\delta(\bbox{r} - \bbox{r'}) -
V^{-1}\right) = \nu \sum_{\alpha \ne 0} \phi_{\alpha}
(\bbox{r}) \phi_{\alpha}(\bbox{r'}).
\end{eqnarray*}
Here $\nu$ is the density of states,  $\phi_{\alpha}(\bbox{r})$ is the
eigenfunction of the Laplace operator with eigenvalue $-E_{\alpha}$;
$\phi_0 = V^{-1/2}$ is the so-called zero-mode. In addition, the
function $K$ contains also a random part. The latter gives rise to the
fluctuations of the charging energy  and is for a moment
ignored. The bare Coulomb potential $U_0$ can be expanded as follows:
\begin{eqnarray} \label{Cul03}
& & U_0(\bbox{r} - \bbox{r'}) = \sum_{\alpha \ne 0} (4\pi
e^2/\epsilon) E_{\alpha}^{-1} \phi_{\alpha}(\bbox{r})
\phi_{\alpha}(\bbox{r'}) 
\nonumber \\
& & + V^{-1/2} \sum_{\alpha \ne 0} u_{0\alpha}
\left[\phi_{\alpha}(\bbox{r}) + 
\phi_{\alpha}(\bbox{r'}) \right] + V^{-1}u_{00},  
\end{eqnarray} 
where we have denoted
\begin{eqnarray} \label{not}
& & u_{0\alpha} = \frac{e^2}{\epsilon V^{1/2}} \int d\bbox{r}
\phi_{\alpha}(\bbox{r}) 
\Phi(\bbox{r}), \ \ \ u_{00} = \frac{e^2}{\epsilon V} \int d\bbox{r}
\Phi(\bbox{r}), \nonumber \\
& & \qquad \Phi(\bbox{r}) = \int d \bbox{r'} \vert
\bbox{r} - \bbox{r'} \vert^{-1}.
\end{eqnarray}
Then the solution to the equation (\ref{rpa1}) reads as
\begin{eqnarray} \label{Cul3}
& & U(\bbox{r},\bbox{r'}) = \sum_{\alpha \ne 0} (4\pi
e^2/\epsilon) (E_{\alpha} + \kappa^2)^{-1} \phi_{\alpha}(\bbox{r})
\phi_{\alpha}(\bbox{r'}) \nonumber \\
& & + V^{-1/2} \sum_{\alpha \ne 0} u_{0\alpha} E_{\alpha} (E_{\alpha} +
\kappa^2)^{-1} \left[\phi_{\alpha}(\bbox{r}) + 
\phi_{\alpha}(\bbox{r'}) \right] \nonumber \\
& & + V^{-1} (u_{00} - \nu \sum_{\alpha \ne 0} (1 +
\kappa^2/E_{\alpha})^{-1} u_{0\alpha}^2),
\end{eqnarray} 
with $\kappa = \sqrt{4\pi e^2 \nu/\epsilon}$ being the inverse
screening length. One can show \cite{BMM} that in the limit $\kappa R
\gg 1$ ($R$ is characteristic size of the system), which we assume
from now on, RPA result (\ref{Cul3}) can be also obtained from the
Thomas-Fermi approximation (TFA), that assumes proportionality
between effective potential and excess charge density.   

Up to a constant, the first term in Eq. (\ref{Cul3}) is the usual 3D
screened Coulomb interaction,
$$U_{\kappa} (\bbox{r_1}, \bbox{r_2}) = (e^2/\epsilon r) \exp(-\kappa
r), $$ 
where $r = \vert \bbox{r_1} - \bbox{r_2} \vert$. Two
other terms appear due to the restricted geometry. The last one is a
constant, $e^2/C$, and $C$ is usual electrostatic capacitance of the
system; for the sphere with radius $R$ we recover $C=\epsilon R$.
Relative correction to $C$ due to the finite value of screening parameter
$\kappa$ is found to be of the order $(\kappa R)^{-1}$. Finally, the
second term (to be denoted as $\tilde U(\bbox{r}) + \tilde U(\bbox{r'})$)
is a single-particle (rather than a two-particle) contribution. It
vanishes in the electrostatic limit $\kappa \to \infty$, and for the
sphere takes the explicit form (up to a constant)
\begin{equation} \label{spart}
\tilde U(\bbox{r}) = -(e^2/\epsilon \kappa R^2) \exp (-\kappa (R - r)),
\end{equation}
i.e. it is localized in a narrow layer of thickness $\kappa^{-1}$ near
the boundary. Thus, the Coulomb interaction in the
restricted geometry is given by
\begin{eqnarray} \label{Culf}
& & U(\bbox{r}, \bbox{r'}) = U_{\kappa} (\bbox{r}, \bbox{r'}) + \tilde
U(\bbox{r}) + \tilde U(\bbox{r'}) + e^2/C; \nonumber \\
& & \qquad \int d\bbox{r} U_{\kappa} (\bbox{r}, \bbox{r'}) = \int d\bbox{r}
\tilde U(\bbox{r}) = 0. 
\end{eqnarray}

The form (\ref{Culf}) can be easily explained. The static screening is
the dressing of the added electron by a positive cloud. This cloud is
created by rearrangement of other electrons, and due to the charge
conservation an excess negative charge $e$ is generated. In the infinite
system this excess charge moves to infinity, and does not play any role.
However, in the finite system it can only move to the boundary, and
creates the additional constant potential $e^2/C$ inside the system. The
potential $\tilde U$ is the edge effect: density of the excess charge
grows towards the boundary, which in TFA gives rise to the additional
nonuniform potential $\tilde U$.  Thus, the total potential is the sum of
the contributions from the positive cloud $(U_{\kappa})$ and from the
negative charge pushed out to the boundary $(\tilde U + e^2/C)$.  Note
that the terms $U_{\kappa}$ and $\tilde U$, ignored by CEM, turn out to
be important for the peak spacing fluctuations.

This consideration can be generalized to the 2D geometry. The function
$K$ in Eq. (\ref{rpa1}) has a form ($\bbox{r} = (x,y)$)
$$ K(\bbox{r}, \bbox{r'}, z, z') = \nu \sum_{\alpha \ne 0} \phi_{\alpha}
(\bbox{r}) \phi_{\alpha}(\bbox{r'}) \delta (z) \delta (z'). $$  
Technically it is convenient to solve this equation in the cylinder, 
and then consider the potential $U(\bbox{r}, \bbox{r'})$ confined
to the plane $z = z' = 0$. The result has the same form (\ref{Culf})
with $\kappa = 2\pi e^2 \nu/\epsilon$, the charging energy 
$$e^2/C = V^{-1} (u_{00} - \nu \sum_{\alpha \ne 0}
(1 + \kappa E_{\alpha}^{-1/2})^{-1} u_{0\alpha}^2),$$
and the 2D screened Coulomb interaction 
$U_{\kappa}(\bbox{r},\bbox{r'})$ 
given by 
\begin{equation} \label{Cul2}
U_{\kappa} (\bbox{r},\bbox{r'}) = \sum_{\alpha \ne 0} (2\pi
e^2/\epsilon) (E^{1/2}_{\alpha} + \kappa)^{-1} \phi_{\alpha}(\bbox{r})
\phi_{\alpha}(\bbox{r'}).
\end{equation}
The single-particle potential $\tilde U(\bbox{r})$ takes the form
\begin{equation} \label{spart2}
\tilde U (\bbox{r}) = V^{-1/2} \sum_{\alpha \ne 0} E_{\alpha}^{1/2}
(E_{\alpha}^{1/2} + \kappa)^{-1}  u_{\alpha 0} \phi_{\alpha} 
(\bbox{r})   
\end{equation}
Here $u_{0\alpha}$ and $u_{00}$ are given by
Eq. (\ref{not}), with the integration going over the 2D
coordinates. In the limit $\kappa R\gg 1$ these results are again
equivalent to TFA. In particular, for a circle with radius $R$ the
potential $\tilde{U}(\bbox{r})$ is
\begin{equation} \label{U2D}
\tilde U(\bbox{r}) = -e^2(2\kappa R)^{-1} (R^2 - r^2)^{-1/2}. 
\end{equation}
Note that, in contrast to the 3D case,  $\tilde U(\bbox{r})$
is a smooth function of magnitude $\sim e^2/\kappa R^2$ all over the
sample. 

In the further consideration we replace the bare Coulomb potential
$U_0(\bbox{r_1} - \bbox{r_2})$ in Eq.~(\ref{int2}) by the screened
potential $U(\bbox{r_1, r_2})$ given by Eq.~(\ref{Culf}). 

{\bf Peak positions}. The position of the conductance peak $\mu_N$
in the Coulomb blockade regime is given by the variation of the total
energy of the system $E(N)$ when an additional (excess) electron is added:
$\mu_N = E(N+1) - E(N)$; the distance between two adjacent peaks
is the difference $S_N = \mu_{N+1} - \mu_N$.

We use Hartree-Fock (HF) approximation and account for $e-e$ interaction
by introducing the effective single-particle Hamiltonian $\hat{H}_N$
which depends explicitly on the number of excess electrons $N$. Its
spectrum is schematically shown on Fig.~1.  First empty state is
separated from the last filled one by a gap of order $e^2/C$. The key
point for the further discussion is that the Hamiltonian $\hat H_N$ is
essentially random for each $N$. Namely, we assume that the statistical
properties of its {\em single-particle} excited states (with the
exception of the gap mentioned above) are the same as that of single
electron states in a random potential. In particular, in the leading
approximation they obey RMT, an assumption which is in agreement with the
experiment \cite{Sivan}.

\vspace{-0.3cm}
\begin{figure}
\centerline{\psfig{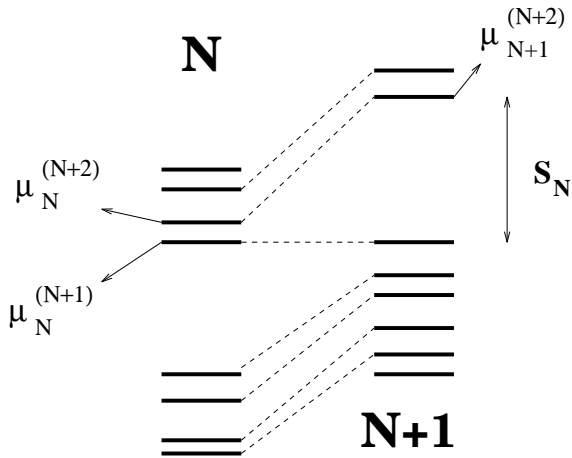}}
\vspace{0.3cm}
\caption{Energy states of the HF Hamiltonian with $N$ (left)
and $N+1$ (right) excess electrons.} 
\label{fig1}
\end{figure}

The distance between two adjacent peaks can be decomposed in a following
way (Fig.~1):
\begin{eqnarray*}
S_N = (\mu_{N+1}^{(N+2)} - \mu_{N}^{(N+2)}) +
(\mu_{N}^{(N+2)} - \mu_{N}^{(N+1)}) \equiv E_1 + E_2.
\end{eqnarray*} 
Here $\mu_i^{(j)}$ is the energy of single-particle eigenstate $j$ of the
(HF) Hamiltonian $\hat H_i$.

Both quantities $E_1$ and $E_2$ are random. The statistical properties
of the latter one are trivial, since it is just a distance between two
adjacent single-particle levels of {\em the same} random Hamiltonian
$\hat{H}_N$. Hence, they obey RMT; in particular, the average $\langle
E_2 \rangle = \Delta$, while the fluctuations are given by
Eq. (\ref{rmt}). 

On the other hand, the quantity $E_1$ is the shift of the $N+2$-th
single-particle level due to addition of a new electron in the $N+1$-th
state. It can be expressed as
\begin{eqnarray} \label{dE1}
E_1 = \int d\bbox{r_1} d\bbox{r_2} U(\bbox{r_1},\bbox{r_2})
\left[ \vert \psi_{N+1} (\bbox{r_1}) \psi_{N+2} (\bbox{r_2}) \vert^2
\right. 
\nonumber \\
- \left. \psi^*_{N+1} (\bbox{r_1}) \psi_{N+1} (\bbox{r_2}) \psi_{N+2} 
(\bbox{r_1}) \psi^*_{N+2} (\bbox{r_2}) \right].
\end{eqnarray}
Here $\psi_{N+1}$ and $\psi_{N+2}$ are eigenfunctions of two first
excited single-particle states of the Hamiltonian $H_N$ (with
corresponding energies $\mu_N^{(N+1)}$ and $\mu_N^{(N+2)}$,
respectively; see Fig.~1). The quantities $E_1$ and $E_2$ are
statistically independent; therefore fluctuations of $S_N$ are given
by the sum of fluctuations of $E_1$ and $E_2$. In other words, since
$E_1$ is related to the change in the Hamiltonian (from $\hat{H}_N$
to $\hat{H}_{N+1}$),  it leads to fluctuations,
{\em additional} to those given by RMT. 

Statistical properties of this quantity can be investigated with the
use of the expressions for correlations of eigenfunctions \cite{BM1}:
\begin{eqnarray} \label{fin2}
& & V^2\langle \vert \psi_k(\bbox{ r}_1) \psi_l(\bbox{ r}_2) \vert^2
\rangle - 1 = k_d(r)[\delta_{kl} + \Pi(\bbox{r}_1,\bbox{r}_1)]
\nonumber \\
& & + \Pi
(\bbox{r}_1, \bbox{r}_2) \delta_{kl} + (1/2) \Pi^2 (\bbox{r}_1,
\bbox{r}_2).
\end{eqnarray}
The short-range function $k_d(r)$ is defined as
\begin{eqnarray*}
& & k_d(r) = (\pi\nu)^{-2}\left[ \mbox{Im} G^R(\bbox{ r})\right]^2 = \\
& & \simeq \exp (-r/l) \left\{
\begin{array}{ll} 
J_0^2(p_F r), & \ \ 2D\\
(p_Fr)^{-2} \sin^2 p_Fr, & \ \ 3D 
\end{array}
\right.
\end{eqnarray*}
and $\Pi(\bbox{r}_1,\bbox{ r}_2)$ is the diffusion propagator. For $r
\gg l$ and for the case of spherical (3D) and circular (2D) particle
\begin{eqnarray} \label{pi1}
& & \Pi(\bbox{r}_1,\bbox{r}_2) 
\approx \left\{
\begin{array}{ll}
(\pi g)^{-1} \ln R/r, & \qquad 2D\\
2R/(3gr), & \qquad 3D 
\end{array}
\right. ,
\end{eqnarray}
with $g = 2\pi E_c/\Delta$ being the dimensionless conductance; $E_c =
D/R^2$ and $D$ are the Thouless energy and the diffusion coefficient,
respectively. For a moment we assume electrons to be spinless, then the
wave functions $\psi_{N+1}$ and $\psi_{N+2}$ are different. 

It can be  shown \cite{BMM}, that the average $\langle E_1
\rangle =
e^2/C$, with negligible corrections of the order $\Delta (p_F l)^{1-d}$.
Thus, the average spacing between the Coulomb blockade
peaks is $\langle S_N \rangle = e^2/C + \Delta$, in agreement with
CEM. 

Now we turn to fluctuations of the quantity $E_1$. In
accordance with decomposition (\ref{Culf}), there are three
contributions. First, the charging energy $e^2/C$ is a
fluctuating quantity due to fluctuations in the polarization operator
 $K(\bbox{r}, \bbox{r'})$ \cite{BA}. A direct calculation \cite{BMM}
shows $\mbox{r.m.s}\ (e^2/C) \sim \alpha_d^2 \Delta \ln g/g$. Here
the parameters $\alpha_3 = 4\pi e^2\nu/\epsilon \kappa^2$ and
$\alpha_2 = 2\pi e^2\nu/\epsilon \kappa$ are equal to unity for the
case of weak interaction ($\kappa \ll p_F$ -- see below). 

Next, we evaluate the fluctuations due to the screened interaction
$U_{\kappa}$. Expanding the average of eight 
wave functions in the cumulants \cite{Bl1} and using Eq.(\ref{fin2})
we find that these 
fluctuations are of order $\mbox{r.m.s}\ [ E_1 (U_{\kappa})]
\sim \alpha_d g^{-1} \Delta$.
Finally, the fluctuations due to the potential $\tilde U$ can be
directly evaluated with the use of Eq. (\ref{fin2}), yielding
\begin{eqnarray} 
\label{corr2}
&& \mbox{r.m.s}\ [E_1 (\tilde U)] \nonumber \\
&&=[2\int
d\bbox{r_1}d\bbox{r_2}\tilde{U}(\bbox{r_1})\tilde{U}(\bbox{r_2})
\langle|\psi_{N+1}^2(\bbox{r_1})\psi_{N+1}^2(\bbox{r_2})|\rangle
]^{1/2} \nonumber \\
&& \sim \left \{ \begin{array}{lr} 
\alpha_2 g^{-1/2} \Delta, & 2D, \\
\alpha_3  g^{-1/2} \Delta, & 3D,
\end{array} \right. 
\end{eqnarray}
which constitutes the main contribution to the fluctuations of the
quantity $E_1$. This result reflects the fact that fluctuations of the
density of the last added electron lead to fluctuations of
its energy in the non-uniform potential $\tilde U(\bbox{r})$.

In the above treatment the screened Coulomb potential has been
calculated within RPA, and thus we have assumed $\kappa \ll p_F$, that
implies 
$\alpha_d = 1$. If the interaction is strong (but still the
ground state of the dot is Fermi-liquid-like), the range of the
screened potential is determined by the Fermi wavelength. The
quantity $\alpha_d$ is then of order of 
$\alpha_d \sim e^2/\epsilon v_F >1$. The limitation on $\alpha_d$
given by the Wigner crystallization is $\alpha < \alpha_c$, where
$\alpha_c \simeq 100$ in 2D (see e.g. \cite{Ando}). Therefore, even
for $\alpha$ well below the Wigner crystallization threshold, the
fluctuations (\ref{corr2}) can exceed $\Delta$, if the dimensionless
conductance $g$ is not too large. 

Although the above results are derived for diffusive
systems ($R \gg l$), we believe also that they are valid for
chaotic ballistic systems ($R < l$) as well. For the latter case the
parameter $g$ can be roughly estimated as the ratio of inverse time of
flight $t_f$ to the level spacing, $g \sim (t_f \Delta)^{-1} \sim
v_F/(R \Delta)$, up to a geometry-dependent coefficient. 

{\bf Spin effects}. We denote as $\uparrow$
and $\downarrow$ two states with the same energy but different values
of spin; their eigenfunctions are identical. If the state $\uparrow$ is
occupied by an electron, the energy of the state $\downarrow$ is
shifted, the shift being  
\begin{equation} \label{spin}
E_1 = \int d\bbox{r_1} d\bbox{r_2} U_{\kappa} (\bbox{r_1},
\bbox{r_2}) \vert \psi_{\lambda} (\bbox{r_1}) \vert^2 \vert
\psi_{\lambda} (\bbox{r_2}) \vert^2.
\end{equation} 
Using Eq. (\ref{fin2}), we find
\begin{equation} \label{spin1}
\langle E_1 \rangle  = \left\{ \begin{array}{lr}
c_d \Delta
(\kappa/p_f)^{d-1} \ln (p_F/\kappa), & e^2/\epsilon \ll v_F  \\
\sim \alpha_d \Delta, & e^2/\epsilon \gtrsim v_F 
\end{array} \right. ,
\end{equation}
with the coefficients $c_3 = 1/2$ and $c_2 = 1/\pi$. 
In the weak interaction regime
($e^2/\epsilon \ll v_F$) we obtain $\langle E_1 \rangle
\ll \Delta$.  
This means that the set of peaks is split into pairs; the reduced
spacing $S_N - e^2/C$ between two peaks of the same pair (states
$\uparrow$ and $\downarrow$) is equal to $\langle E_1 \rangle$,
while that 
between two different pairs is of order $\Delta$. Furthermore, since the
eigenfunctions of states $\uparrow$ and $\downarrow$ are identical,
the corresponding peak heights are correlated, and the resulting
picture is a set of pairs with small reduced spacings and correlated
heights -- a pronounced odd-even structure. Fluctuations of
$E_1$ are given by Eq. (\ref{corr2}) multiplied by a factor of 2,
and do not destroy the pair structure. 

For strong interaction ($e^2/\epsilon \gtrsim v_F$) the splitting 
(\ref{spin1}) exceeds $\Delta$, and reordering of energy levels 
takes place. However, the spacing fluctuations are still enhanced in
comparison with the spinless value, since a new type of  
ensemble is created,
with the spectrum  formed by a superposition of a
RMT-type set of levels and the same set shifted by $E_1$, 
$$\nu(E) = \sum_{\lambda} \left[ \delta(E - E_{\lambda}) + \delta(E +
E_1 - E_{\lambda}) \right].$$
In particular, the two-point
correlation function is given by ($\Delta$ is now the mean level
spacing for a system with removed spin degeneracy, $s = \pi \omega/
2\Delta$, $\tilde s = \pi E_1 / 2\Delta$) 
\begin{eqnarray} \label{spin2}
& & R(\omega) \equiv \langle \nu(E) \nu(E + \omega) \rangle / \langle
\nu(E) \rangle \langle \nu(E + \omega) \rangle \nonumber \\ 
& & = \frac{1}{2} R_{RMT} (s) + \frac{1}{4}
R_{RMT} (s - \tilde s) + \frac{1}{4} R_{RMT} (s
+ \tilde s), 
\end{eqnarray}
where  $R_{RMT} (s) = 1 - s^{-2} \sin^2 s$ is the RMT
two-point correlation function. The level repulsion in the
ensemble (\ref{spin2}) is clearly reduced. Thus,
fluctuations of peak spacings are enhanced by spin
effects, in addition to their enhancement due to fluctuations in
$E_1$ discussed above, Eq.(\ref{corr2}). 
On the other hand, even-odd correlations of heights of
adjacent peaks in the strong interaction regime are washed out. 

In conclusion, we have studied Coulomb interaction effects on the 
statistics of conductance peaks spacings in the Coulomb blockade
regime. The Coulomb interaction leads, in combination with
fluctuations of eigenfunctions, to enhancement of
peak spacing fluctuations in comparison with the RMT value.
Taking into account the spin degrees of freedom, we find in addition
in the case of weak 
interaction ($e^2/\epsilon \ll v_F$) a 
pronounced odd-even structure. The peaks come in pairs, with correlated
heights and small reduced spacings $S - e^2/C \ll \Delta$ within
each pair. In the opposite case $e^2/\epsilon \gtrsim v_F$, which is
relevant to the experiment \cite{Marcus,Sivan1}, this structure is
destroyed. 

We are grateful to L.I.~Glazman and C.M.~Marcus for useful
discussions. This work was supported by the Alexander von Humboldt
Foundation (Y.M.B.), SFB 195 der Deutschen Forschungsgemeinschaft
(Y.M.B., A.D.M.), and the German-Israeli Foundation (A.D.M.).


\begin{thebibliography}{99}

\vspace{-1.cm}

\item[$^*$] Present address: D\'epartement
de Physique Th\'eorique, Universit\'e de Gen\`eve,  
CH-1211 Gen\`eve 4, Switzerland.

\bibitem{Marcus} J.~A.~Folk et al, 
Phys. Rev. Lett. {\bf 76}, 1699 (1996); C.~M.~Marcus et al, 
{\em Superlattices and Microstructures}, ICSMM, Liege
(1996) (to be published); C.~M.~Marcus (private communication). 

\bibitem{Sivan1} U.~Sivan et al, 
Phys. Rev. Lett. {\bf 77}, 1123 (1996). 

\bibitem{CIT} See e.g. M.~A.~Kastner, Rev. Mod. Phys. {\bf 64}, 437
(1992).  

\bibitem{Mehta} M.~L.~Mehta, {\em Random Matrices}, Academic Press
(N.Y.) (1991). 

\bibitem{Prus} O.~Prus et al, Phys. Rev. B {\bf 54}, R14289 (1996). 

\bibitem{BM1} Ya.~M.~Blanter and A.~D.~Mirlin, Phys. Rev. B {\bf
53}, 12601 (1996); Phys. Rev. E {\bf 55}, May (1997).

\bibitem{Buett} M.~B\"uttiker, J. Phys. Cond. Mat. {\bf 5}, 9361 (1993).

\bibitem{BA} R.~Berkovits and B.~L.~Altshuler, Phys. Rev. B {\bf 46},
12526 (1992). 

\bibitem{BMM} Ya.~M.~Blanter, A.~D.~Mirlin, and B.~A.~Muzykantskii
(unpublished). 

\bibitem{Sivan} U.~Sivan et al, 
Europhys. Lett. {\bf 25}, 605 (1994).

\bibitem{Bl1} Ya.~M.~Blanter, Phys. Rev. B {\bf 54}, 12807 (1996). 

\bibitem{Ando} T.~Ando, A.~B.~Fowler, and F.~Stern,
Rev. Mod. Phys. {\bf 54}, 437 (1982).

\end{thebibliography}
\end{document}